\documentclass[usenatbib]{mnras}

\usepackage[T1]{fontenc}
\usepackage{ae,aecompl}

\usepackage{graphicx}	% Including figure files
\usepackage{amsmath}	% Advanced maths commands
\usepackage{amssymb}	% Extra maths symbols
\usepackage{mathrsfs}
\usepackage{enumitem}
\usepackage[normalem]{ulem} % Editing
\usepackage{multirow} % Merging table cells
\usepackage{xcolor} % Highlighting
\usepackage{lipsum} % Lorem Ipsum
\usepackage{hyperref} % Eprint and doi bibtex fields shall be hyperlinks
\usepackage{mathtools} % For \coloneqq command etc.
\usepackage{subcaption}
\usepackage{array} 
\usepackage{lipsum}
\usepackage{txfonts}
\usepackage{float}
\usepackage{physics}
\usepackage[shortcuts]{extdash}
\usepackage{lipsum}  
\usepackage{dblfloatfix}
\usepackage{siunitx}
\usepackage{booktabs}
\usepackage{amstext} % for \text macro
\usepackage{array}   % for \newcolumntype macro
\newcolumntype{L}[1]{>{\centering\arraybackslash$}p{#1}<{$}}
\usepackage{pifont}% http://ctan.org/pkg/pifont
\newcommand{\cmark}{\ding{51}}%
\newcommand{\xmark}{\ding{55}}%
\DeclareRobustCommand{\VAN}[3]{#2}
\let\VANthebibliography\thebibliography
\def\thebibliography{\DeclareRobustCommand{\VAN}[3]{##3}\VANthebibliography}

\makeatletter
\def\footnoterule{\kern-3\p@
  \hrule \@width 2in \kern 2.6\p@} % the \hrule is .4pt high
\makeatother

\usepackage{scalerel,tikz}
\usetikzlibrary{svg.path}
\definecolor{orcidlogocol}{HTML}{A6CE39}
\tikzset{orcidlogo/.pic={
 \fill[orcidlogocol] svg{M256,128c0,70.7-57.3,128-128,128C57.3,256,0,198.7,0,128C0,57.3,57.3,0,128,0C198.7,0,256,57.3,256,128z};
 \fill[white] svg{M86.3,186.2H70.9V79.1h15.4v48.4V186.2z}
 svg{M108.9,79.1h41.6c39.6,0,57,28.3,57,53.6c0,27.5-21.5,53.6-56.8,53.6h-41.8V79.1z M124.3,172.4h24.5c34.9,0,42.9-26.5,42.9-39.7c0-21.5-13.7-39.7-43.7-39.7h-23.7V172.4z}
 svg{M88.7,56.8c0,5.5-4.5,10.1-10.1,10.1c-5.6,0-10.1-4.6-10.1-10.1c0-5.6,4.5-10.1,10.1-10.1C84.2,46.7,88.7,51.3,88.7,56.8z};
}}
\newcommand\orcidicon[1]{\href{https://orcid.org/#1}{\mbox{\scalerel*{
\begin{tikzpicture}[yscale=-1,transform shape]
\pic{orcidlogo};
\end{tikzpicture}
}{|}}}}

\title[Increased Burstiness in Multi-Physics Models]{Increased Burstiness at High Redshift in Multi-Physics Models Combining Supernova Feedback, Radiative Transfer and Cosmic Rays}

\author[T. Dome et al.]
{\parbox[t]{\textwidth}
{Tibor Dome$^{1,2}$\thanks{E-mail: td448@cam.ac.uk}, 
Sergio Martin-Alvarez$^{1,4}$,
Sandro Tacchella$^{2,3}$,
Yuxuan Yuan$^{1,2}$~\orcidicon{0000-0001-6816-0682}, and
Debora Sijacki$^{1,2}$
}
\\ \\
$^{1}$Institute of Astronomy, University of Cambridge, Madingley Road, Cambridge, CB3 0HA, UK\\
$^{2}$Kavli Institute for Cosmology, Madingley Road, Cambridge, CB3 0HA, UK\\
$^{3}$Cavendish Laboratory, University of Cambridge, 19 JJ Thomson Avenue, Cambridge, CB3 OHE, UK\\
$^{4}$Kavli Institute for Particle Astrophysics and Cosmology (KIPAC), Stanford University, Stanford CA 94305, USA
}

\date{MNRAS, submitted}

\pubyear{2024}

\begin{document}
\label{firstpage}
\pagerange{\pageref{firstpage}--\pageref{lastpage}}
\maketitle

\begin{abstract}
We study star formation variability, or burstiness, as a method to constrain and compare different galaxy formation models at high redshift using the \textsc{Azahar} simulation suite. The models range from magneto-hydrodynamics with a magneto-thermo-turbulent prescription for star formation (iMHD) to more sophisticated setups incorporating radiative transfer (RTiMHD) and cosmic ray physics (RTnsCRiMHD). Analysing a sample of galaxies at redshifts $z=4-10$, we find that the RTnsCRiMHD model exhibits more regular star formation periodicity compared to iMHD and RTiMHD, as revealed by the Lomb-Scargle periodogram. While the RTiMHD model captures a notable degree of stochasticity in star formation without cosmic rays, RTnsCRiMHD galaxies display even greater scatter in the burst intensity and in the scatter around the star-forming main sequence. To evaluate the burstiness in RTnsCRiMHD against observations, we generate a mock spectrum during a mini-quenching event at $z=7.5$. This spectrum aligns well with the low-mass quiescent galaxy JADES-GS-z7-01-QU observed at $z=7.3$, though some discrepancies attributed to stellar metallicity hint at a composite spectrum. Our findings highlight the importance of including complex physical processes like cosmic rays and radiative transfer in simulations to accurately capture the bursty nature of star formation in high-redshift galaxies. Future JWST observations, particularly regarding the scatter around the star-forming main sequence, have the potential to refine and guide the next generation of galaxy formation models.\end{abstract}


%%%%%%%%%%%%%%%%%%%%%%%%%%%%%%%%%%%%%%%%%%%%%%%%%%

%%%%%%%%%%%%%%%%% BODY OF PAPER %%%%%%%%%%%%%%%%%%

\section{Introduction}
\label{s_intro}
The first quiescent galaxies offer a unique probe of the mechanisms behind the cessation of star formation \citep{Schreiber_2018, Girelli_2019, Merlin_2019, Nanayakkara_2024, Carnall_2023_b, Long_2023, Valentino_2023, Carnall_2024}. These galaxies formed in the early universe, when the available Hubble time to assemble and subsequently quench galaxies was particularly short. Spectroscopic studies suggest a fast build-up of stellar mass up to $10^{11} \ M_{\odot}$ within the first one or two billion years of the Universe, followed by rapid quenching within a few tens of million years \citep{Carnall_2023, Kakimoto_2024, Glazebrook_2024}, possibly challenging our current understanding of galaxy formation \citep{Lovell_2023, Finkelstein_2023}.\par

The majority of early quiescent galaxies identified to date are massive systems ($M_{\star} > 10^{10} \ M_{\odot}$), yet the \textit{James Webb Space Telescope} (JWST) is opening a new window on the first generations of faint low-mass galaxies populating the high-redshift Universe. \cite{Looser_2024} reported the discovery of JADES-GS-z7-01-QU, a $M_{\star} = 10^{8.6} \ M_{\odot}$ quiescent galaxy at a reionisation-epoch redshift of $z=7.3$, while \cite{Strait_2023} reported the quiescent galaxy MACS0417-z5BBG with an even lower stellar mass (for the bulge of the galaxy) of $M_{\star} = 10^{7.6} \ M_{\odot}$ at $z=5.2$. The first spectroscopic confirmation of galaxies that have rejuvenated following a quiescent phase was reported by \cite{Witten_2024b}, with mass $M_{\star} = 10^{9.6} \ M_{\odot}$ at $z=7.9$. These low-mass quiescent galaxies might have quenched their star formation through different physical processes than their higher-mass counterparts.\par 

The observability of galaxies near the limiting flux of a survey (typically bursty low-mass and/or high-redshift galaxies) is highly time-dependent due to star formation rate (SFR) variability \citep{Sun_2023}. Due to its weak Balmer break, JADES-GS-z7-01-QU would not be identified as passive using the traditional rest-frame UVJ diagram \citep{Williams_2009}. To identify predominantly massive quiescent galaxies at high redshift, some studies propose a modified UVJ selection \citep{Belli_2019}, while others favour a NUVrJ \citep{Ilbert_2013}, FUVVJ \citep{Leja_2019} or ugi color selection \citep{Danso_2023}. For lower-mass systems, \cite{Trussler_2024} propose a photometric search method for `smouldering' galaxies based on medium-band NIRCam imaging to identify the lack of emission lines. High-redshift low-mass quiescent galaxies similar to JADES-GS-z7-01-QU have also been found at $z \approx 4.5$ \citep{Looser_2023b}. The peculiar nebular emission displayed by early galaxies with Balmer line ratios inconsistent with Case B recombination \citep{Yanagisawa_2024, Pirzkal_2024} might also hint at quenching events, specifically a density bounded transient phase of $\sim 20$ Myr following a rapid dissipation of H \textsc{ii} regions \citep{McClymont_2024}.\par

The observations by \cite{Looser_2024} are in rough agreement with the semi-analytical model GAEA, which suggests that the first quenched low-mass galaxy ($M_{\star}\lesssim 10^{9.5} \ M_{\odot}$) is expected to appear before $z=7$ \citep{Xie_2024}. Since many (mostly massive) high-redshift quenched galaxies are found to host luminous active galactic nuclei \citep[AGN,][]{Belli_2024, Davies_2024, Eugenio_2023}, feedback from their central supermassive black holes might constitute an important contribution to quenching. \textsc{IllustrisTNG}, \textsc{Magneticum} and \textsc{Flares} simulations at $z \sim 3$ indeed indicate that massive galaxies are typically quenched by AGN feedback \citep{Lovell_2022, Kurinchi_2023, Kimmig_2023}. The recent study by \cite{Xie_2024} suggests that disc instabilities can trigger efficient black hole accretion and subsequent quenching via (quasar) winds also for lower-mass systems ($M_{\star} \approx 10^{9.5} \ M_{\odot}$). At even lower masses of $M_{\star} \approx 10^{7}-10^{9} \ M_{\odot}$, overmassive black holes in the centers of dwarfs may give rise to efficient AGN feedback at high redshift without violating observed H~\textsc{i} gas mass constraints \citep{Koudmani_2022}, though observational evidence of such feedback channels is missing. `Mass quenching' such as virial shock heating or the loss or removal of gas from a galaxy due to ram pressure may contribute at the high-mass end as early as $z\sim 4$ \citep{Tanaka_2023, Alberts_2023}, though this is less likely for low-mass galaxies at $z\gtrsim 5$.\par 

In this paper, we aim to make progress on the modelling front and focus on the physical mechanisms that might give rise to JADES-GS-z7-01-QU-like spectra. Since both JADES-GS-z7-01-QU and MACS0417-z5BBG appear to be in the post-starburst phase, the quenching might have been driven by internal (feedback) processes. Numerical simulations underscore the necessity for feedback processes beyond supernovae (SNe) alone \citep[e.g.,][]{Smith2019, Alvarez_2023}, with simplified physical models \citep{Gelli_2024} suggesting that these do not suffice to quench galaxies of $M_{\star} \approx 10^8 \ M_{\odot}$ at high redshift on timescales of $\approx 30$ Myr required by JADES-GS-z7-01-QU. But if not conventional SN feedback, what other internal or external feedback processes can explain JADES-GS-z7-01-QU? In \cite{Dome_2024}, we showed that many quenched galaxies in \textsc{IllustrisTNG} \citep{Pillepich_2018, Pillepich_2019, Nelson_2019} are gravitationally interacting with other galaxies (more than $25$\% at $z = 7$), and even when not fully merging are tidally disrupted \citep[see also][]{Asada_2024}. However, in \cite{Dome_2024} we could only reconcile mock spectral energy distributions (SEDs) extracted from \textsc{IllustrisTNG}, \textsc{Vela} \citep{Ceverino_2014, Zolotov_2015}, and \textsc{FirstLight} \citep{Ceverino_2017} with JADES-GS-z7-01-QU spectrophotometry when artificially modifying the ages of stellar populations, revealing the missing baryonic physics in these simulations.\par 

\cite{Faisst_2024} suggest that dust distributed along the line of sight could attenuate nebular emission and mimic quiescence. They find that spatially varying dust attenuation (assuming a blue+red composite spectrum) can reproduce a JADES-GS-z7-01-QU-like spectrum with a blue UV to optical $\lambda f_{\lambda}$ flux ratio and lack of emission lines, even though the red component may not be quiescent. Strong stochastic variations in the SFH can facilitate this effect, with $60-80$ per cent of all galaxies in the high-resolution \textsc{Sphinx} cosmological radiation hydrodynamic simulations \citep{Rosdahl_2018, Rosdahl_2022, Katz_2023b} exhibiting similar observed properties (lack of emission lines, blue UV continuum) to JADES-GS-z7-01-QU at some point during their lifetime above $z = 7$.\par

In this work, we use SFR variability as a tool to constrain numerical models of galaxy formation. Our approach involves quantifying SFR variability across a range of models with increasing physical complexity and identifying episodes of temporary quiescence \citep[a.k.a. mini-quenching,][]{Dome_2024} that most accurately reproduce the spectrophotometry of JADES-GS-z7-01-QU. We review the ability of different models to reproduce these observations by investigating a subset of simulations of the upcoming \textsc{Azahar} suite of cosmological zoom-in simulations of galaxy formation (Martin-Alvarez et al. in prep). \textsc{Azahar} begins with fiducial star formation and stellar feedback models and progressively incorporates more sophisticated physics, culminating in `multi-physics' simulations featuring magneto-thermo-turbulent (MTT) star formation, mechanical SN feedback with magnetic fields, on-the-fly radiative transfer, and cosmic rays. A key advantage of using \textsc{Azahar} for the purpose of constraining numerical models is that model parameters are selected purely from physical considerations, without any aim to match specific observables through parameter calibration.\par 

The organisation of the paper is as follows. A brief description of the \textsc{Azahar} simulations and the different physical models is provided in Sec.~\ref{ss_tng_sims}. Details on how we calculate stellar masses and star-formation rates can be found in Secs.~\ref{ss_av_timescale} and \ref{ss_aperture}. Methods to generate mock SEDs are described in Sec.~\ref{ss_seds}. We discuss our main findings in \mbox{Sec.~\ref{s_conclusions}}.

\renewcommand{\arraystretch}{1.6}
\begin{table*}
	\centering
	\caption{Galaxy formation models: (1) Name; (2) and (3) whether the simulation accounts for radiative transfer and cosmic rays, respectively; (4) the stellar feedback modelling; (5) further details regarding the characteristics of the simulation; (6) and (7) the halo and stellar mass of the secondary galaxy (G2) at $z=7$, respectively. Note that all galaxy formation models adopt $B_0 = 3\times 10^{-20}$~G as the initial, uniform magnetic field seed and the MTT prescription for star formation (see text).} 
	\noindent\begin{tabular}{p{1.6cm}p{0.8cm}p{0.8cm}p{3cm}p{5cm}p{1.5cm}p{1.5cm}}
	\hline
    Name & RT & CR & Stellar feedback & Further details & $M_{\mathrm{h}}(z=7)$ & $M_{\star}(z=7)$\\
	\hline
	iMHD & \xmark & \xmark & MagMech & SN inject $E_{\text{mag,SN}}$ & $6.4\times 10^{11} \ M_{\odot}$ & $3.2 \times 10^{9} \ M_{\odot}$\\
	RTiMHD & \cmark & \xmark & Radiation + MagMech & RT magnetism; SN inject $E_{\text{mag,SN}}$ & $2.3\times 10^{11} \ M_{\odot}$ & $1.6 \times 10^{9} \ M_{\odot}$\\
	RTnsCRiMHD & \cmark & \cmark & Radiation + CRMagMech & RT magnetism; SN inject $E_{\text{mag,SN}}+E_{\text{CR,SN}}$ & $2.2\times 10^{11} \ M_{\odot}$ & $7.9 \times 10^{8} \ M_{\odot}$\\
	\bottomrule
	\end{tabular}
	\label{t_sims}
\end{table*}
\renewcommand{\arraystretch}{1}
\section{Theoretical Models and Post-Processing}
\label{s_num_methods}

\subsection{\textsc{Azahar} Simulations}
\label{ss_tng_sims}
In order to study quenching in simulations that incorporate many physical processes, we take advantage of the \textsc{Azahar} simulation suite, briefly described here and to be presented in detail by Martin-Alvarez et al. (in prep). The \textsc{Azahar} simulations are a comprehensive set of models that integrate hydrodynamics, magnetic fields, radiative transfer, and cosmic ray physics to study their effects and interactions in galaxy formation. These simulations use the magneto-hydrodynamical code RAMSES \citep{Teyssier_2002}, which ensures divergence-free magnetic field evolution through a constrained transport method \citep{Teyssier_2006}.\par
\begin{figure}
\hspace{-0.45cm}
\includegraphics[width=0.52\textwidth]{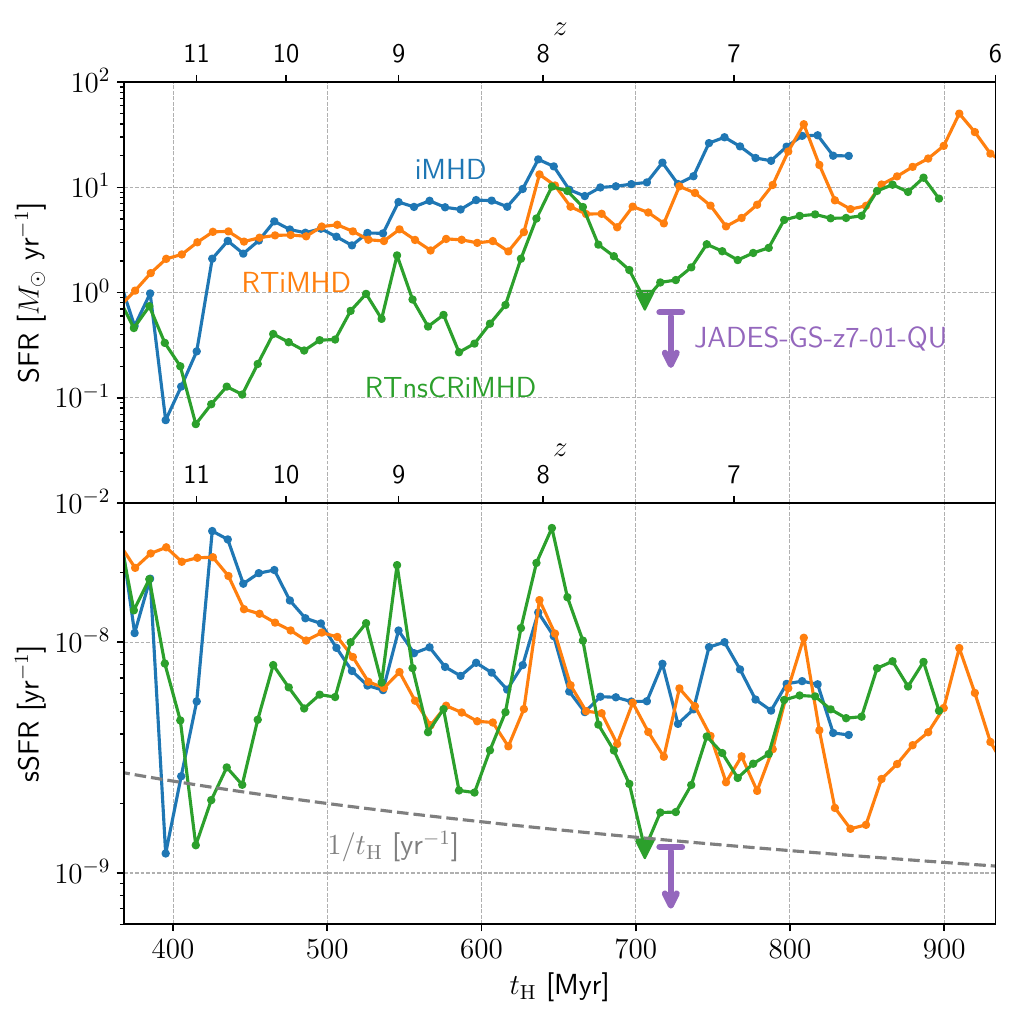}
\caption{\textsc{Azahar} SFHs in three galaxy formation models: iMHD, RTiMHD and RTnsCRiMHD. We show the SFR (top panel) and sSFR (bottom panel) histories of the secondary galaxy G2 before merging with the primary galaxy G1 at $z\lesssim 6.5$. The mini-quenching event at $z=7.5$ of the RTnsCRiMHD G2 galaxy is highlighted with a downward-pointing triangle marker and the corresponding SED is presented in Fig.~\ref{f_sed}. The upper limit on the (s)SFR of JADES-GS-z7-01-QU is also shown (purple), along with the inverse Hubble time, $1/t_{\mathrm{H}}$, indicated by a dashed grey line in the bottom panel. Note that RTnsCRiMHD galaxies are visibly more bursty than their iMHD and RTiMHD counterparts, which we quantify in Secs.~\ref{s_period} and \ref{s_quantburst} for a sample of galaxies.}
\label{f_sfh}
\end{figure} 

Within its extensive zoom region ($8$ cMpc across), \textsc{Azahar} features two main progenitors for its most massive system at high redshift ($z \gtrsim 6$). The primary and secondary galaxies are identified in this work as G1 and G2, respectively. These galaxies form within the zoom region, and merge at $z \lesssim 6.5$ to eventually reach a halo mass of around $2.5 \times 10^{12} \ M_{\odot}$ at $z=1$. The maximum spatial resolution achieved in these simulations is $\Delta x \approx 20$ pc (full cell-width), with refinements triggered based on mass criteria as well as Jeans length, fundamental for resolving ISM turbulence and magnetic fields \citep{Alvarez_2022}. Dark matter particles in the zoom region have a mass of $m_{\text{DM}} \approx 4.5 \times 10^5 \ M_{\odot}$, and stellar mass particles have a mass of $m_{\star} \approx 4 \times 10^4 \ M_{\odot}$. The suite builds on the physics framework established by the pathfinder simulation, Pandora \citep{Alvarez_2023}.\par

In this study, we focus on iMHD, RTiMHD and one of the ``multi-physics'' simulations in the \textsc{Azahar} suite, RTnsCRiMHD. The RTnsCRiMHD model was already employed in \citet{Witten_2024}. All of these simulations include detailed treatments of radiative cooling above and below $10^4$ K \citep{Rosen_1995, Ferland_1998}, star formation following an MTT prescription \citep{Federrath_2012, Kimm_2017,Alvarez_2020}, and mechanical supernova (SN) feedback\footnote{\cite{Zhang_2024} recently showed that varying only the directional distribution of momentum imparted from SNe to the surrounding gas can already amplify or completely suppress bursty star formation, and affect the total stellar mass formed by as much as a factor of $\approx 3$.} \citep{Kimm_2014}. MTT allows for a spatially varying gas-to-star conversion efficiency parameter (the star formation efficiency $\varepsilon$), replacing the commonly used constant $\varepsilon$ model which spawns stars once the gas density exceeds a threshold value with a fixed efficiency $\varepsilon \approx 0.015$. We deem this necessary for a high-redshift star formation prescription not least because SN events in MTT take place later than in the constant $\varepsilon$ model, in regions of higher physical density, leading to star formation being burstier with MTT \citep[see Pandora framework,][]{Alvarez_2023}. In addition, the ultra-massive quiescent galaxies identified by \cite{Carnall_2024} roughly align with the most massive galaxies expected in $\Lambda$CDM under the assumption of $100$ per cent conversion of baryons to stars ($\varepsilon=1$), further motivating sub-grid models that go beyond the simple constant $\varepsilon$ models.\par 

Additionally, all the studied models include magnetic fields, with a negligible seed primordial field, $B_0 = 3\times 10^{-20}$ G, and primarily sourced by magnetised SN feedback. This magnetised feedback injects 1\% of SN energy as magnetic energy ($E_{\text{mag,SN}}=0.01 E_{\text{SN}} \sim 10^{49}$ erg) to simulate SN remnant magnetisation, and is described in further detail by \citet{Alvarez_2021}. The RTiMHD model includes on-the-fly radiative transfer \citep{Rosdahl_2015} with a configuration similar to that of the \textsc{Sphinx} simulations. It employs three energy bins for radiation covering different ionisation energy intervals: from $13.6$ eV to $24.59$ eV (H~\textsc{i} ionisation), $24.59$ eV to $54.42$ eV (He~\textsc{i} ionisation), and above $54.42$ eV (He~\textsc{ii} ionisation). Finally, for RTnsCRiMHD, the simulation also incorporates cosmic rays, modelled as an anisotropically diffusing energy density \citep{Dubois_2016, Dubois_2019}, with 10\% of supernova energy converted into cosmic rays, $E_{\text{CR,SN}}=0.1 E_{\text{SN}} \sim 10^{50}$ erg. The cosmic ray diffusion coefficient is set to $\kappa_{\text{CR}} = 3 \times 10^{28}  \ \text{cm}^2 \text{s}^{-1}$, without accounting for cosmic ray streaming. We summarise the key assumptions of the three galaxy formation models in Table~\ref{t_sims}. Further details on the numerics and model configurations are provided in the Pandora pathfinders \citet{Alvarez_2023}.\par 

To illustrate a typical SFH in \textsc{Azahar}, we follow the evolution of the galaxy G2 across the three galaxy formation models: iMHD, RTiMHD and RTnsCRiMHD. Fig.~\ref{f_sfh} shows their SFHs in terms of the SFR averaged over $t_{\text{avg}} = 10$ Myr. In addition, we show the specific SFR, defined as:
\begin{equation}
\mathrm{sSFR} = \frac{\mathrm{SFR}}{M_{\star}}.
\end{equation}
While all models exhibit a generally bursty mode of star formation at the redshifts shown ($z\geq 4$), the RTnsCRiMHD model shows a markedly higher level of burstiness (i.e., greater stochastic fluctuations) compared to iMHD and RTiMHD. In the RTnsCRiMHD model, sSFR values can fluctuate by more than $1$ dex over short timescales of just a few tens of Myrs, around $\approx 30$ Myr. While both the \textsc{Sphinx} and \mbox{\textsc{Fire}-2} simulations also show high levels of burstiness \citep{Katz_2023b,Sun_2023b}, possibly sufficient to account for the abundance of bright galaxies at cosmic dawn, our focus here is on the difference in stochasticity across different galaxy formation models. In contrast, the behaviour observed in iMHD aligns with our previous findings \citep{Dome_2024}, where galaxies in the \textsc{IllustrisTNG}, \textsc{Vela} and \textsc{FirstLight} simulations typically do not exhibit this degree of burstiness; in these models, a $1$ dex change in sSFR unfolds over longer timescales, exceeding \mbox{$\approx 50$ Myr}.\par 

\subsection{Simulated Galaxy Sample}
\label{ss_gx_sample}
For most of this study, we focus on the redshift range $z=4-10$ and select galaxies with stellar masses $M_{\star} = 10^8-10^{10} \ M_{\sun}$ at $z=4$ from the preliminary \textsc{Azahar} galaxy catalogues. This yields around $10$ galaxies per simulated model. Among these, the galaxies G1 and G2 are included in the sample. Halo properties are computed using {\sc HaloMaker} \citep{Tweed_2009} on the dark matter particles.\par 

For the preliminary \textsc{Azahar} catalogues, all halos with $M_\text{halo} > 10^{8} \ M_\odot$ at $z = 10$ are seeded with a galaxy tracker. Each tracker follows a single galaxy using its (up to) 500 innermost stellar particles. These catalogues only contain and follow galaxies already formed at $z = 10$, and therefore exclude the least massive galaxies and those that formed after $z = 10$. This new {\sc ramses} galaxy catalogue pipeline will be introduced and publicly released by Martin-Alvarez et al. (in prep). More details are provided by \citet{Alvarez_2024a} for its offline version and \citet{Sanati_2024} for its preliminary on-the-fly version.

\subsection{Averaging Timescales}
\label{ss_av_timescale}
Simulations and observations consistently show that shorter SFR averaging timescales result in a higher normalisation and increased scatter in the star-forming main sequence \citep[MS,][]{Speagle_2014, Popesso_2023}, particularly at the low-mass end \citep{Schaerer_2013, Hayward_2014, Speagle_2014, Sparre_2015, Caplar_2019, Donnari_2019, Tacchella_2020}. This phenomenon is partly due to observational sampling biases that favour young stars or regions of intense star formation. In our study, the minimum averaging timescale is $t_{\text{avg}} = 10$~Myr, which is essential for accurately resolving the star formation histories of galaxies and studying star formation stochasticity. For timescales $t_{\text{avg}} > 10$~Myr, SFRs are derived by integrating SFH$_{10}$ (SFR averaged over $10$~Myr) and then normalising by the integration period. 

\subsection{Aperture Choice}
\label{ss_aperture}
The stellar mass and SFR of a galaxy are estimated based on the formation times and masses of star particles evaluated between consecutive snapshots. These estimates are influenced by the radius within which they are measured. To ensure consistency with common methodologies, we adopt a standard approach as outlined by \citet{Alvarez_2018} or \citet{Sun_2023}. Specifically, we define the stellar mass $M_{\star}$, gas mass $M_{\text{gas}}$, SFR, and gas metallicity $Z_{\text{gas}}$ as the summed (or projected) contributions from stellar particles and gas cells within $0.2 \ R_{\text{vir}}$ of the halo center, where $R_{\text{vir}}$ denotes the virial radius \citep{Bryan_1998}. Changing the aperture does not significantly affect the results (see Sec.~\ref{ss_caveats}).

\subsection{Calculating Spectral Energy Distributions}
\label{ss_seds}
To compute a dust-free, nebular emission-free spectral energy distribution (SED) of a galaxy, we treat each stellar particle in \textsc{Azahar} as a simple stellar population (SSP) using a stellar population synthesis method. Specifically, we use the Flexible Stellar Population Synthesis (FSPS) code \citep{Conroy_2009, ConroyGunn_2010} with MIST isochrones \citep{Paxton_2011, Paxton_2013, Paxton_2015, Choi_2016, Dotter_2016} and the MILES stellar library \citep{Sanchez_2006, Falcon_2011}. We assume a Kroupa initial mass function \citep{Kroupa_2001}, consistent with the sub-grid model assumptions used in \textsc{Azahar}.\par 

To account for dust attenuation and extinction, we follow the methodology of \cite{Nelson_2018}. This model, when applied to \textsc{IllustrisTNG} simulations, accurately replicates the observed distribution of optical $(g-r)$ colours from the Sloan Digital Sky Survey (SDSS). We employ a simple power-law extinction model \citep{Charlot_2000} to account for attenuation by finite-lifetime birth clouds around young stellar populations and the ambient diffuse ISM. In addition, we track the distribution of metals and neutral hydrogen gas within and around each simulated galaxy. Neutral hydrogen fraction estimates for each gas cell are derived from the \textsc{Azahar} output, which captures a multi-phase interstellar medium, and self-consistently models hydrogen ionisation in the simulations including radiative transfer (i.e., RTiMHD and RTnsCRiMHD models).\par 

The \cite{Nelson_2018} resolved dust model then assigns a neighbourhood- and viewing angle-dependent attenuation to each stellar particle. Specifically, the absorption optical depth reads
\begin{equation}
\tau_{\lambda}^a=\left(\frac{A_{\lambda}}{A_{\text{V}}}\right)_{\odot}(1+z)^{\beta}\left(\frac{Z_{\text{gas}}}{Z_{\odot}}\right)^{\gamma}\frac{N_{\text{HI}}}{N_{\text{HI,0}}},
\end{equation}
where the first term is the solar neighbourhood extinction curve \citep{Cardelli_1989}, while the second and third terms parametrise a redshift and metallicity-dependent dust-to-gas ratio \citep[see also][]{McKinnon_2016}. The values $Z_{\text{gas}}$ and $N_{\text{HI}}$ depend on the projected position of the stellar particle and are determined through interpolation on the gas projection grids.\footnote{The projection grids for gas metallicity and neutral hydrogen gas have a pixel scale of $15$ pc.} We adopt $\beta = -0.5$ and a broken power law ($\gamma = 1.6$ for $\lambda > 200$~nm and $\gamma = 1.35$ for $\lambda < 200$~nm, outside the SDSS bands). We take the solar metallicity as $Z_{\odot} = 0.0127$ and for the normalisation of the hydrogen column we adopt $N_{\text{HI,0}} = 2.1 \times 10^{21}$~cm$^{-2}$. We will later discuss the impact of different viewing angles, selected as vertices of the $N_s = 1$ \mbox{\scshape{HEALPIX}} \normalfont sphere \citep{Gorski_2005} aligned with simulation coordinates.\par

To model the nebular emission associated with a stellar population (both emission lines and continuum emission), we use the photoionisation code CLOUDY \citep{Chatzikos_2023}. For a given low-density birth cloud exposed to ionising radiation (i.e. an H \textsc{ii} region), CLOUDY determines the thermal, ionisation, and chemical state of the cloud and the resultant spectrum of the transmitted radiation. While a simple model of CLOUDY is integrated into FSPS as CLOUDYFSPS \citep{Byler_2017}, we run CLOUDY directly for each young stellar particle (with age $t< 10$~Myr) sitting at the centre of a birth cloud and acting as its ionising source. We choose $t=10$~Myr as the threshold since nebular emission is negligible compared to the stellar spectrum after around $10$~Myr \citep{Zackrisson_2011}, when the birth cloud has usually been disrupted or stars have migrated away.\par 

For the photoionisation modelling choices we follow the approach of \cite{Charlot_2001} and \cite{Wilkins_2020} who characterise the brightness of the incident radiation field using the ionisation parameter at the Strömgren radius, $R_{\text{S}}$. The ionisation parameter is defined as the ratio of hydrogen-ionising photon to hydrogen densities,
\begin{equation}
U_{\text{S}} = U(R_{\text{S}}) = \frac{\alpha_{\text{B}}^{2/3}}{3c}\left(\frac{3Q\kappa^2n_{\text{H}}}{4\pi}\right)^{1/3}.
\label{e_U_at_stromgren}
\end{equation}
Here, $n_{\text{H}}$ is the total hydrogen density including ionized, neutral, and molecular hydrogen, $c$ is the speed of light, and
\begin{equation}
Q=M_{\star}\int_{13.6 \ \text{eV}}^{\infty} \frac{L_{\nu}}{h\nu} \mathrm{d}\nu,
\end{equation}
is the rate of ionising photons emitted by the source per second, where the monochromatic luminosity, $L_{\nu}$, is taken directly from the intrinsic SED generated by FSPS with nebular emission turned off. The volume-filling factor of the gas is denoted as $\kappa$, and $\alpha_{\text{B}}$ is the case-B hydrogen recombination coefficient \citep{Osterbrock_2006}. Note that Eq.~\eqref{e_U_at_stromgren} assumes that the inner radius of the gaseous nebula $r_{\text{in}}$ is significantly smaller than $R_{\text{S}}$, $r_{\text{in}} \ll R_{\text{S}}$, and neglects the weak dependence of $\alpha_{\text{B}}$ on $r$ through the electron temperature.\par

Since the vast majority of H~\textsc{ii} regions are only marginally resolved in \textsc{Azahar}, we follow \cite{Wilkins_2020} and express the ionisation parameter $U_{\text{S}}$ relative to a reference value $U_{\text{S,ref}}$, defined at a reference age ($t = 1$ Myr) and metallicity ($Z = 0.02$). The actual ionisation parameter passed to CLOUDY thus depends on the ionising photon production rate relative to the reference value,
\begin{equation}
U_{\text{S}}=U_{\text{S,ref}}\left(\frac{Q}{Q_{\text{ref}}}\right)^{1/3}.
\end{equation}
This fixes the assumed geometry of the H~\textsc{ii} region, encoded in the $\kappa^2 n_{\text{H}}$ term, for different stellar metallicities and ages. We assume $\log_{10}(U_{\text{S,ref}})=-2$ and $\log_{10}(n_{\text{H}}/\text{cm}^{-3})=-2.5$. The metallicity of the gas cloud is set to match that of the corresponding stellar particle. For the depletion of metals onto dust grains in the gas cloud, we adopt Orion-type graphite and silicate grains which can boost certain lines and provide an additional source of attenuation \citep{Nakajima_2018}. The interstellar abundances of metals and dust depletion factors are taken from \cite{Gutkin_2016}. We adopt the default CLOUDY stopping temperature ($4000$ K), which is suitable for UV/optical recombination lines.\par

\section{Periodicity Analysis}
\label{s_period}
To study the theoretical characteristics of SFHs across different galaxy formation models, we quantify the amplitude and timescales of the star formation stochasticity following \citet{Pallottini_2023}, who examined high-redshift galaxies in the \textsc{Serra} radiation hydrodynamics simulations \citep{Pallottini_2022}. For each galaxy in the sample detailed in Sec.~\ref{ss_gx_sample}, we fit the average SFR trend using a simple polynomial in logarithmic space:
\begin{equation}
\log_{10} \langle \mathrm{SFR}/M_{\sun} \ \mathrm{yr}^{-1}\rangle = \sum_{i = 0}^{2}p_i\left(\frac{t_{\text{H}}}{\mathrm{Myr}}\right)^i.
\label{e_sfms}
\end{equation}
The polynomial is limited to second order to minimise the influence of potential oscillatory terms in the SFR. Typically, the best-fit coefficients satisfy $p_1>0$ and $p_2 <0$, indicating that the SFR for most galaxies begins at a low level, rises to a peak, and then declines. This behaviour aligns with the delayed exponentially-declining SFR model, or time-delayed SFR, commonly adopted in semi-analytical models of galaxy formation \citep{Chiosi_2017}. This pattern can also be understood through extended Press-Schechter-based toy models of galaxy formation, which suggest that a galaxy's SFR tends to reach a steady state following the accretion rate onto the galaxy \citep{Bouche_2010, Lilly_2013, Dekel_2013_b, Tacchella_2018}.\par 

We then define the stochastic time variability of the star formation as the residual of the fit:
\begin{equation}
\delta = \log_{10}\frac{\mathrm{SFR}}{\langle\mathrm{SFR}\rangle}.
\end{equation}
For each galaxy, $\delta$ is approximately distributed as a zero-mean Gaussian, with maximum amplitudes reaching around $|\delta_{\text{max}}|\approx 1.0$ and in rare cases, particularly for RTnsCRiMHD, as high as $|\delta_{\text{max}}|\approx 1.5$. These amplitudes are significantly higher than found for typical \textsc{Serra} galaxies \citep[$|\delta_{\text{max}}|\approx 0.7-0.8$,][]{Pallottini_2023}, hinting at particularly strong bursts and quiescent phases in RTnsCRiMHD.\par 
\begin{figure}
\hspace{-0.45cm}
\includegraphics[width=0.52\textwidth]{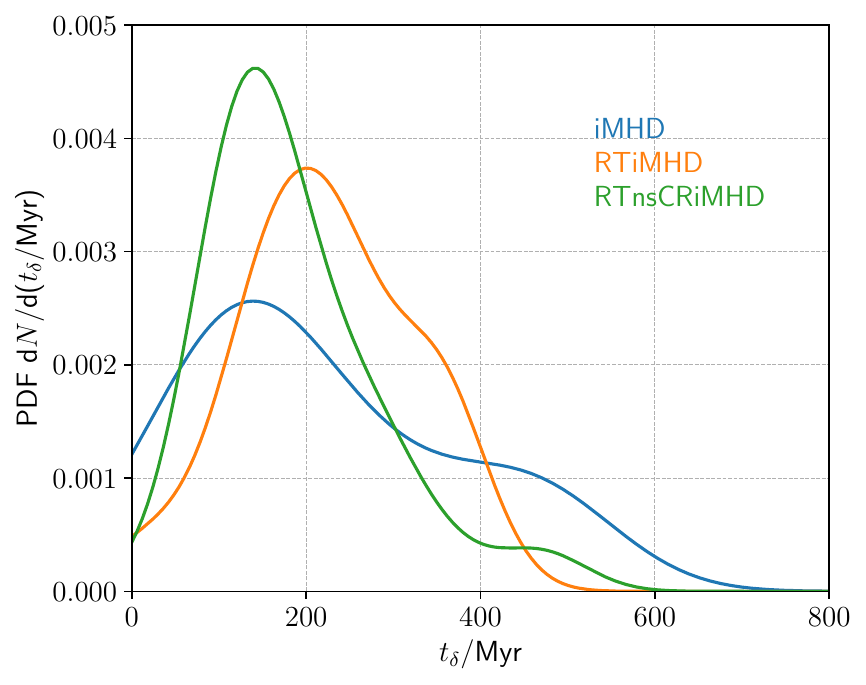}
\caption{Distribution of the characteristic \protect\cite{Lomb_1976}-\protect\cite{Scargle_1998} timescales $t_{\delta}$ of the SFH periodicity in three galaxy formation models of \textsc{Azahar}: iMHD, RTiMHD and the comprehensive multi-physics model RTnsCRiMHD. The PDF is computed via a kernel density estimator, adopting \protect\cite{Silverman_1986}'s rule-of-thumb estimate for the bandwidth size and weighting by the peak significance $w$. Only peaks above the noise threshold of $P(t_{\delta}) = 0.04$ are considered. Note that the peak of the distribution is significantly more pronounced (i.e., greater height and narrower width) in the RTnsCRiMHD model than in the iMHD and RTiMHD models.}
\label{f_periodogram}
\end{figure}

To analyse periodicity in the time series $\delta(t)$ for each galaxy, we employ the \cite{Lomb_1976}-\cite{Scargle_1998} periodogram\footnote{For an accessible introduction, see \cite{VanderPlas_2018}, who explains that the Lomb-Scargle method is not only rooted in Fourier analysis and least-squares fitting, but can also be derived from Bayesian probability theory. Moreover, it shares similarities with bin-based phase-folding techniques under certain conditions.}. One of the practical advantages of the Lomb-Scargle method is its ability to clearly identify peaks and conveniently assess their significance using the false alarm probability, defined as $w = 1 - \mathrm{false} \ \mathrm{alarm}$. In contrast, traditional power spectrum estimation, given by PSD$(k)=|f(k)|^2$, where $f(k)$ is the Fourier transform of an evenly sampled SFH, $f(k) = 1/\sqrt{\Delta t}\int \mathrm{d}te^{-ikt}\psi(t)$, often struggles to isolate individual frequency contributions and lacks a straightforward method to quantify peak significance \citep{Iyer_2020, Shin_2022}. Using the Lomb-Scargle periodogram, we analyse each galaxy's time series, selecting all peaks $t_{\delta}$ above the noise threshold of $P(t_{\delta}) = 0.04$. We then compute the probability density function (PDF) of the timescales for the sample by weighting each $t_{\delta}$ with its \mbox{significance $w$}.\par 

The resulting distribution of characteristic timescales $t_{\delta}$ is shown in Fig.~\ref{f_periodogram}. We observe a pronounced peak at a characteristic timescale of $t_{\delta}\sim 100-200$~Myr in each of the three galaxy formation models. This peak corresponds to a modulation consistent with cosmological accretion/merging timescales of massive ($M_{\mathrm{h}} \sim 10^{11}\ M_{\odot}$) dark matter halos at redshifts $z\simeq 6 - 10$ \citep{Furlanetto_2017}, but is also influenced by feedback strength, outflows, and gas depletion. In addition, this timescale is comparable to the dynamical timescale ($t_{\text{dyn}}\approx 1/\sqrt{G\rho}$) of intermediate-mass galaxies at redshifts $z=4-10$, suggesting a significant interplay between these processes.\par 

The peak is accompanied by a long tail extending up to $t_{\delta} \gtrsim 400$~Myr in all models. This tail is notably more extended than the one found by \cite{Pallottini_2023}, which we attribute to our longer averaging timescale of $\Delta t = 10$~Myr. A larger $\Delta t$ acts like a smoothing filter, averaging out short-term fluctuations and highlighting longer, smoother trends. In addition, with a lower sampling rate, there is a risk of aliasing, where higher-frequency signals are misinterpreted as lower-frequency signals. Due to our stellar mass resolution of $m_{\star} \approx 4 \times 10^4 \ M_{\odot}$ (see Sec.~\ref{ss_tng_sims}), reducing $\Delta t$ significantly would result in Poisson sampling issues \citep[e.g.][]{Iyer_2020}.\par

Importantly, the peak in the $t_{\delta}$ distribution is significantly more pronounced (i.e., greater height and narrower width) in the comprehensive multi-physics model RTnsCRiMHD than in the iMHD and RTiMHD models. The regularity of the modulation is thus substantially enhanced in RTnsCRiMHD, suggesting that the inclusion of cosmic rays and radiative transfer not only leads to bursty SFHs (as evidenced by the width of the $\eta$ distribution) but also promotes episodic starbursts interspersed with quiescent phases at more regular intervals. The enhanced periodicity observed in RTnsCRiMHD is likely induced, at least partially, by the driving of more significant galactic outflows in the presence of cosmic rays \citep[e.g.,][]{Girichidis2018, Hopkins2021, Rodriguez_Montero_2024}. This enhanced periodicity underscores the role of complex physical processes in shaping the star formation dynamics within galaxies.\par 

\section{Increased Burstiness in Multi-Physics Models}
\label{s_quantburst}
While the periodicity analysis provides insights into the timescales that shape typical SFHs, the periodogram of a galaxy cannot be directly observed. Therefore, we quantify the level of burstiness in \textsc{Azahar} using practical measures to better understand the physical mechanisms driving the stochastic variability in the SFHs of simulated galaxies. Burstiness, in this context, refers to episodic bursts of intense star formation activity followed by periods of relative quiescence, where a galaxy forms significantly fewer stars over a short timescale ($\sim 10$~Myr) compared to its longer-term average ($\sim 100$~Myr). During ``regular'' phases near the star-forming MS, galaxies form stars at comparable rates over both short and long timescales.\par 
\begin{figure*}
\begin{subfigure}{0.495\textwidth}
\includegraphics[width=\textwidth]{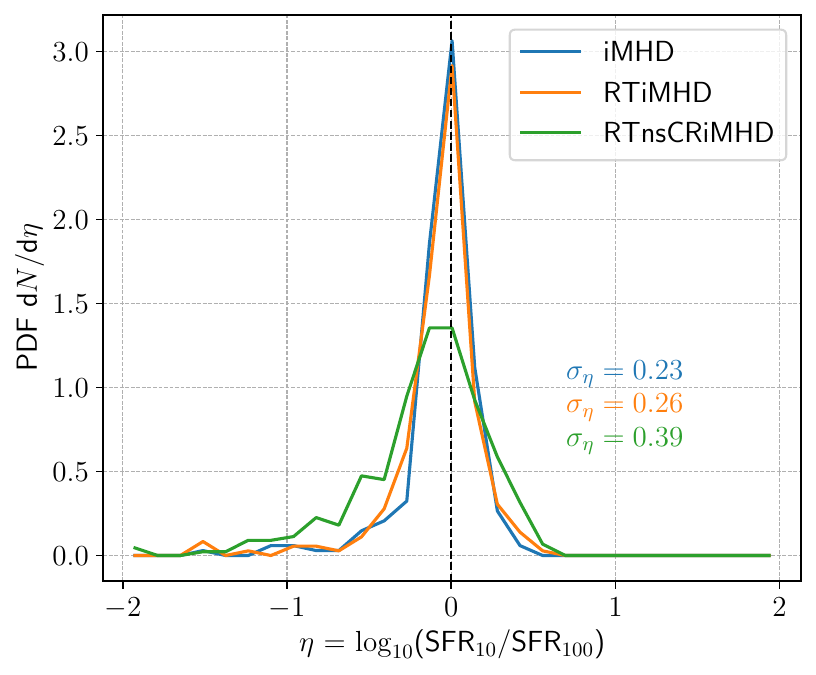}
\end{subfigure}
\begin{subfigure}{0.495\textwidth}
\includegraphics[width=\textwidth]{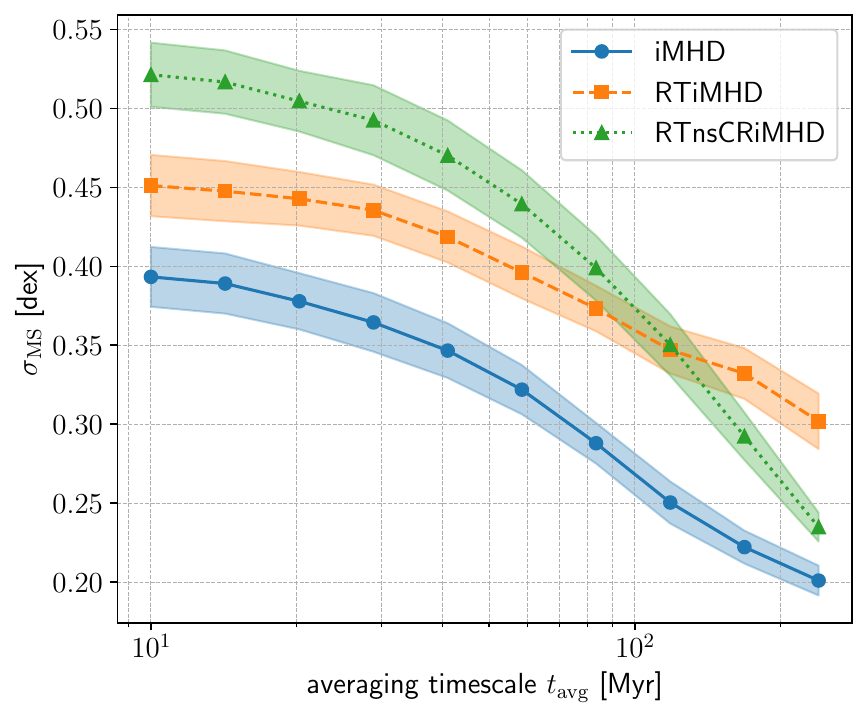}
\end{subfigure}
\caption{Burstiness of galaxies in \textsc{Azahar}. \textbf{Left panel}: Distribution of the burst intensity $\eta$ in three galaxy formation models of \textsc{Azahar}: iMHD (blue), RTiMHD (orange) and the comprehensive multi-physics model RTnsCRiMHD (green). The standard deviation of the RTnsCRiMHD galaxy sample is $\sigma_{\eta} = 0.39$~dex, which is notably higher than the values found in the iMHD and RTiMHD models, where $\sigma_{\eta} = 0.23$~dex and $\sigma_{\eta} = 0.26$~dex, respectively. This larger standard deviation in the RTnsCRiMHD model is indicative of more bursty SFHs. \textbf{Right panel}: Scatter around the star-forming MS for galaxies in \textsc{Azahar}. The scatter is measured as the standard deviation of the logarithm of the SFRs, normalised to their position on the star-forming MS according to Eq.~\eqref{e_sfms}. We present the scatter, $\sigma_{\text{MS}}$, as a function of the time-scale $t_{\text{avg}}$ over which the SFR is averaged. The shaded region denotes the standard error obtained through bootstrap resampling. Galaxies in RTnsCRiMHD are more bursty than their iMHD and RTiMHD counterparts, with $\sigma_{\text{MS}}$ reaching approximately $0.52$ at short averaging timescales ($t_{\text{avg}} = 10$~Myr), in contrast to $\sigma_{\text{MS}} \approx 0.39$~dex and $\sigma_{\text{MS}} \approx 0.45$~dex for the iMHD and RTiMHD galaxies, respectively.}
\label{f_scatter}
\end{figure*} 

Consistent with observational work \citep{Weisz_2012,Guo_2016, Broussard_2019, Cole_2023, Looser_2023b, Boyett_2024, Simmonds_2024} and following previous authors using (radiation) hydrodynamics simulations \citep{Dansac_2004, Caplar_2019, Iyer_2020, Katz_2023}, we thus average the newly formed stellar mass over two time windows of widths $10$~Myr (SFR$_{10}$) and $100$~Myr (SFR$_{100}$), respectively. The logarithmic ratio 
\begin{equation}
\eta = \log_{10}(\text{SFR}_{10}/\text{SFR}_{100})
\end{equation}
is sometimes referred to as the burst intensity or burst indicator. \cite{Broussard_2019} showed that the average of $\eta$ is not a reliable measure of burstiness, as it tends to be close to zero unless the galaxy population has a significantly rising or falling average SFH. Indeed, \cite{Boyett_2024} confirmed that galaxies which exhibit large equivalent widths (EW) in their rest-optical emission lines ([OIII]$\lambda$5007 or H$\alpha$ rest-frame EW$>750 \ \AA$) called extreme emission line galaxies can be tied to a recent upturn in their SFR, leading to high burst intensities which can surpass SFR$_{10}$/SFR$_{100} \approx 8$. In contrast, the width of the $\eta$ distribution provides a comprehensive measure of burstiness in a galaxy population's recent star formation. \cite{Broussard_2019} found that this width is relatively stable across different stellar initial mass functions, metallicities, and dust measurement errors, given certain assumptions.\par 

Note that SFR$_{100}$ traces the average star-formation activity over the same timescales as empirical tracers based on the rest-frame NUV emission \citep[e.g.][]{Shivaei_2015} while SFR$_{10}$ traces the rest-frame FUV continuum \citep{Kennicutt_1998}. The EW of H$\alpha$ probes timescales of $3-10$ Myr and is an indirect measure of the ionising continuum, hence observational SFR estimates based on H$\alpha$ are often in good agreement with SFR$_{10}$ estimates inferred from the UV continuum \citep{Looser_2023b}.\par 

In Fig.~\ref{f_scatter} (left panel), we show the distribution of the burst intensity $\eta$ across three \textsc{Azahar} galaxy formation models. We average the newly formed stellar mass over the respective time windows ($10$ and $100$~Myr) at regular intervals spaced $20$~Myr apart. While the resulting SFR estimates cannot all be independent realisations, we assume that the full sample traces the underlying distribution reliably. All models exhibit a prominent peak around $\eta = 0$, suggesting that, on average, the SFR of the galaxy sample is stable, neither increasing nor decreasing. However, the standard deviations of $\eta$ differ: $\sigma_{\eta} = 0.39$~dex for the RTnsCRiMHD model, compared to $\sigma_{\eta} = 0.26$~dex and $\sigma_{\eta} = 0.23$~dex for RTiMHD and iMHD, respectively. This variation in the second moment of the burst intensity suggests that galaxies in the comprehensive multi-physics model RTnsCRiMHD experience burstier SFHs than those in the RTiMHD and iMHD models. Note that the negative skewness in $\eta$ is due to a tendency for galaxies to quench rapidly, dropping to low SFRs ($\eta << 0$), while having no equivalent extreme for starbursts, where a nonzero SFR$_{10}$ necessitates a nonzero SFR$_{100}$ \citep[see][]{Broussard_2019}.\par 

We take the analysis further by quantifying the scatter, $\sigma_{\text{MS}}$, of the star-forming MS as a function of the averaging timescale, $t_{\text{avg}}$. Following the periodicity analysis in Sec.~\ref{s_period}, the star-forming MS is defined for each galaxy individually according to Eq.~\eqref{e_sfms}. Fig.~\ref{f_scatter} (right panel) shows the resulting scatter for timescales $t_{\text{avg}} = 10-250$~Myr. We find that galaxies in the RTnsCRiMHD model exhibit more burstiness than those in the iMHD and RTiMHD models, with $\sigma_{\text{MS}}$ reaching approximately $0.52$ at short averaging timescales ($t_{\text{avg}} = 10$~Myr). In contrast, $\sigma_{\text{MS}}$ levels off at around $0.39$~dex and $0.45$~dex for the iMHD and RTiMHD samples, respectively.\par 

As the SFH decorrelates over longer averaging timescales, $\sigma_{\text{MS}}$ generally declines, reflecting that multiple bursts are being averaged out, making the measured SFR less sensitive to short-term variability. The steepness of this decline is indicative of the timescale over which the SFR remains correlated \citep{Iyer_2022}. The RTnsCRiMHD sample shows the steepest decline around $t_{\text{avg}} = 100$~Myr, while the SFHs in the RTiMHD model remain correlated over slightly longer timescales. However, the behaviour at timescales $t_{\text{avg}} > 150$~Myr should be interpreted with caution, as the limited Hubble time at high redshift might influence these estimates.

\section{Mini-Quenching in \textsc{Azahar}}
\subsection{Bridging the gap with JADES-GS-z7-01-QU}
We now turn our attention to the mini-quenching event of the RTnsCRiMHD G2 galaxy at $z=7.5$. Notably, its sSFR of around $10^{-9}$ yr$^{-1}$ falls within the threshold defined by the inverse Hubble time, $1/t_{\mathrm{H}}$, a common benchmark at intermediate-to-high redshifts to identify quiescent systems \citep{Tacchella_2019, Bluck_2024}. This sSFR is also comparable to the upper limit on the SFR of JADES-GS-z7-01-QU, which is inferred from the upper limit on the H$\beta$ emission-line flux, suggesting an SFR of $\approx 0.65 \ M_{\odot}\mathrm{yr}^{-1}$ and a corresponding sSFR of $\approx 1.3\times 10^{-9}\ \mathrm{yr}^{-1}$ \citep{Looser_2024}. Note that the exact SFR for JADES-GS-z7-01-QU is highly uncertain due to the absence of observed emission lines; hence, we refer to the upper limit.\par 
\begin{figure*}
\begin{subfigure}{0.495\textwidth}
\includegraphics[width=\textwidth]{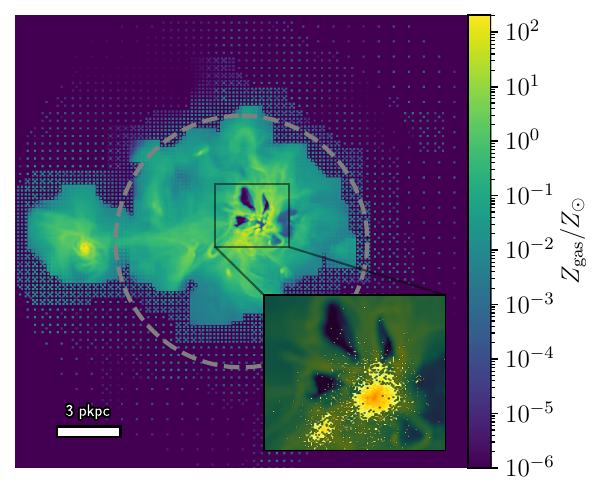}	
\end{subfigure}
\begin{subfigure}{0.495\textwidth}
\includegraphics[width=\textwidth]{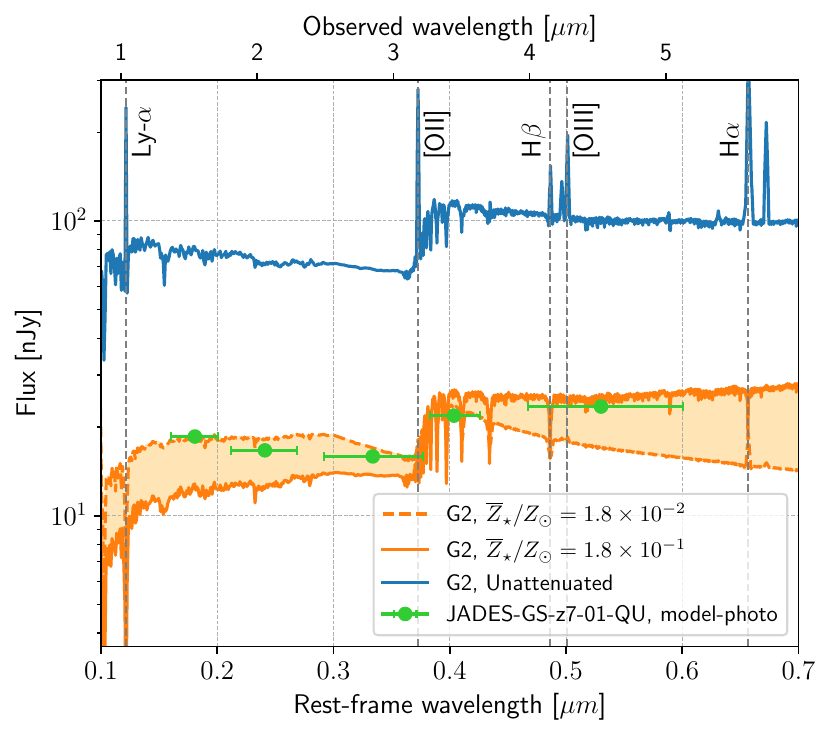}
\end{subfigure}
\caption{\textsc{Azahar} mock spectra compared to JADES-GS-z7-01-QU. \textbf{Left panel}: We show the projection of metals in units of solar metallicity ($Z_{\odot} = 0.0127$) around the secondary galaxy (G2). The inset highlights the spatial distribution of stellar particles, where the colour in the hex-bin plot represents the logarithmic density of stars. A dashed grey circle marks the aperture of $0.2 \ R_{\text{vir}}$. \textbf{Right panel}: We present both the unattenuated spectrum of G2 (blue solid) and the dust-attenuated G2 spectrum (orange solid). \scshape{FORCEPHO} \normalshape model photometry for the main JADES-GS-z7-01-QU galaxy (without the clump) are shown as green dots for filters F150W, F200W, F277W, F335M and F444W, with horizontal bars indicating the wavelength range probed by each passband. Error bars on the flux are not visible at this resolution. The mean stellar metallicity is $\overline{Z}_{\star}/Z_{\odot}=1.8\times 10^{-1}$. To better match the inferred metallicity for JADES-GS-z7-01-QU ($Z/Z_{\odot}\sim 2\times 10^{-2}$), we artificially reduce the metallicity of each stellar particle by $1$~dex (orange dashed), leading to better agreement for filters F150W, F200W and F277W while somewhat deteriorating the agreement for filter F444W. Note that nebular emission lines are clearly visible for the unattenuated spectrum but strongly suppressed for the attenuated spectra. Grey vertical dashed lines trace the strongest lines: Lyman-$\alpha$ ($\lambda = 1216 \ \AA$), [OII]$\lambda$3727, H$\beta$ ($\lambda = 4861 \ \AA$), [OIII]$\lambda$5007 and H$\alpha$ ($\lambda = 6565 \ \AA$).}
\label{f_sed}
\end{figure*}

In Fig.~\ref{f_sed} (right panel), we show the SED of G2 mock-observed at this mini-quenching event calculated as per Sec.~\ref{ss_seds}. To account for differences in redshift and stellar mass between G2 and JADES-GS-z7-01-QU, we apply flux correction factors
\begin{equation}
\frac{d_{\mathrm{L}}^2(7.3)}{d^2_{\mathrm{L}}(7.5)} \approx 0.93,
\end{equation}
where $d_{\mathrm{L}}(z)$ is the luminosity distance at redshift $z$, and 
\begin{equation}
\frac{M_{\star,\text{G2}}}{M_{\star,\text{GS-z7-01-QU}}} \approx 1.34,
\end{equation}
assuming mass-to-light ratio scaling \citep[see][]{Schombert_2019}. Incorporating these correction factors, the unattenuated spectrum of G2 has a flux density that is about a factor of $5$ higher than JADES-GS-z7-01-QU model photometry, with several strong emission lines including Lyman-$\alpha$, [OII]$\lambda$3727, H$\beta$, [OIII]$\lambda$5007 and H$\alpha$.\par 

The attenuated mock SED of G2 (orange solid) is obtained by selecting the optimal line-of-sight using the \cite{Nelson_2018} attenuation model to best match the observed photometry. For the majority of viewing angles, we find significant dust attenuation, yielding fluxes well below those of the observed system. Only about 17\% of the viewing angles show sufficient dust ejection or removal from the emission regions to maintain high flux levels. The optimal line-of-sight successfully brings the overall flux normalisation into close agreement with JADES-GS-z7-01-QU, particularly for filters F335M and F444W, which now align well with the observed photometry. The good match is achievable only with the RTnsCRiMHD model, as similar low sSFR events in iMHD and RTiMHD result in mock SEDs that fall short of observed flux levels \citep[cf.][]{Dome_2024}. This underscores the crucial role of radiation and cosmic rays in reproducing galaxy spectra at redshifts $z=4-10$.\par 

The evacuation of dust-rich gas can be seen in Fig.~\ref{f_sed} (left panel), where stellar feedback has created cavities in the gas metallicity ($Z_{\text{gas}}$) projection (see also Yuan et al., in prep). A similar pattern is observed in the neutral hydrogen column density ($N_{\text{HI}}$, not shown). Given that most stars are older than 10~Myr, clearing the line-of-sight to the emission region does not lead to strong emission lines. In fact, all viewing angles for G2 in the RTnsCRiMHD model experience sufficient dust attenuation to suppress the emission lines from young stellar populations.

\subsection{Stellar Metallicity}
Some differences between the mock SED of G2 and JADES-GS-z7-01-QU remain, most notably in the near-UV (rest-frame), where model photometry fluxes significantly exceed the mock SED. The Balmer break is also significantly stronger for G2 than suggested by the photometry. At first order, both discrepancies can be traced to the high mean stellar metallicity of G2, $\overline{Z}_{\star}/Z_{\odot}=1.8\times 10^{-1}$, which is about 1~dex higher than the inferred metallicity for JADES-GS-z7-01-QU, $Z/Z_{\odot}\sim 2\times 10^{-2}$. Note that the uncertainty on the metallicity inferred from the observations is large since the abundance of metals and dust depletion factors are poorly known at high redshift; \cite{Looser_2024} assumes solar abundances.\par 

By artificially reducing the mean stellar metallicity by 1~dex (orange dashed), we improve the match in the near-UV for filters F150W, F200W and F277W, but at the cost of a poorer match for filter F444W. Note that this problem was already highlighted in \cite{Dome_2024} and is not directly related to bursty star formation in the early Universe. Instead, \cite{Faisst_2024} suggest that this discrepancy hints at a blue+red composite spectrum, where the UV continuum is emitted from dust-free density bounded H~\textsc{ii} regions (blue component), while the red component is a dust-obscured starburst with weakened emission lines due to strong differential dust attenuation. Investigating this further is beyond the scope of this work and not the primary focus. 

\subsection{Caveats}
\label{ss_caveats}
While reducing the metallicity by 1~dex helps align the near-UV filters with observations, it introduces discrepancies elsewhere, hinting at potential composite spectral components that are not fully accounted for in this analysis. Additionally, the effective dust attenuation optical depth in the mock galaxy (RTnsCRiMHD G2) is considerably higher, with $A_{\text{V}}=-2.5\log_{10}(\text{flux ratio}) \ \text{mag}=1.37$~mag, compared to the inferred $A_{\text{V}}<0.57$~mag for the observed system (JADES-GS-z7-01-QU). This upper limit is well-constrained by the observed UV slope and remains robust even when the Lyman-$\alpha$ drop is masked, in which case the fit would indicate a nearly dust-free environment.\par

This discrepancy in dust attenuation can possibly be traced to assumptions made in the forward modelling of the spectrum. Importantly, the \cite{Nelson_2018} model projects gas cell quantities (metallicity and neutral hydrogen) onto an orthographic grid. This flat-screen attenuation model does not take into account the relative position of stars and the absorbing dust in the line-of-sight direction, and likely overestimates dust attenuation. Considering the significantly higher resolution in \textsc{Azahar}, where dust structures are resolved instead of diffuse as in e.g. \textsc{IllustrisTNG}, this is potentially important. However, since we examine a range of viewing angles, a $z$-sorted attenuation model is computationally expensive and beyond the scope of this work. It will be investigated in more detail by Yuan et al. (in prep).\par 

On the other hand, allowing the outflow to develop further may evacuate most of the dust and lead to a reduced value of $A_{\text{V}}$ that better matches the observations, but this intermediate stage is not captured by the output frequency of snapshots. Overall, there is a range of implementation details in dust attenuation modelling to consider. Our results indicate, consistent with our previous work \citep{Dome_2024}, that many differences, such as using H~\textsc{i} versus H column densities, yield only minor variations in the resulting mock SEDs. Similarly, the choice of aperture size has minimal impact on the SED; using twice the stellar half-mass radius instead of $0.2 \ R_{\text{vir}}$ results in approximately a 5\% change in the SED, without significantly affecting the overall shape, absorption features, or emission lines.

\section{Summary \& Outlook}
\label{s_conclusions}
In this work, we employ star formation variability, or burstiness, as a means to constrain and compare different numerical models of galaxy formation at high redshift.\par

\textit{Methods:} We use the \textsc{Azahar} simulation suite, which comprises a range of models. Amongst its models, we select as our most simple model the simulation labelled iMHD, which incorporates hydrodynamics, magnetic fields, a magneto-thermo-turbulent prescription for star formation. We study two additional models that progressively integrate more sophisticated physics: radiative transfer (RTiMHD) and cosmic ray physics (RTnsCRiMHD). The latter represents one of the ``multi-physics'' simulations. We select a small sample of $\approx 10$ galaxies at redshifts $z=4-10$ in each of the three galaxy formation models to analyse the star formation periodicity and quantify the degree of burstiness.

\textit{Periodicity:} We examine the periodicity of star formation using the Lomb-Scargle periodogram \citep{Lomb_1976, Scargle_1998}. The resulting distribution of characteristic timescales $t_{\delta}$ shows a more pronounced peak in the RTnsCRiMHD model. This indicates a more regular periodicity in its SFHs, suggesting that the inclusion of cosmic rays and radiative transfer enhances the regularity of star formation bursts.

\textit{Burstiness:} Moving on to more measurable quantities, the burst intensity indicator $\eta$, which quantifies the ratio of star formation rates over 10~Myr and 100~Myr intervals, shows significantly greater scatter in $\eta$ for the RTnsCRiMHD model. Similarly, the scatter around the star-forming MS, $\sigma_{\text{MS}}$, is greater in RTnsCRiMHD. These results suggest that star formation is significantly more bursty in galaxies modelled with RTnsCRiMHD, highlighting the impact of the included physical processes on star formation behaviour. Since RTiMHD exhibits greater scatter in $\eta$ and higher values of $\sigma_{\text{MS}}$ compared to iMHD, the RTiMHD model still captures a notable degree of stochasticity in star formation without cosmic rays, though not to the extent seen in RTnsCRiMHD.

\textit{Mini-Quenching:} Leveraging on the pronounced burstiness observed in the RTnsCRiMHD model, we generate a forward-modelled mock spectrum during a mini-quenching event at $z=7.5$ and compare it with observational data for the high-redshift quiescent galaxy JADES-GS-z7-01-QU. The resulting mock spectrum shows strong agreement with the \textsc{Forcepho} model photometry for the primary JADES-GS-z7-01-QU galaxy (excluding the clump). However, some residual differences in the stellar metallicity distribution suggest the potential presence of a blue+red composite spectrum, as indicated by \citet{Faisst_2024}.

\textit{Outlook:} This study underscores the critical role of incorporating comprehensive physical processes, such as cosmic rays and radiative transfer, into hydrodynamical simulations to more accurately capture the burstiness and periodicity of star formation in high-redshift galaxies. By quantifying the burstiness of galaxies in \textsc{Azahar} using observable metrics, we bridge the gap between simulations and observations. Future measurements of the scatter around the star-forming main sequence ($\sigma_{\text{MS}}$) using JWST observations, with high completeness, will be pivotal in constraining numerical models of galaxy formation. If $\sigma_{\text{MS}}$ is observed to exceed $0.5$~dex on averaging timescales of $t_{\text{avg}} = 10$ Myr at $z=4-10$, this would provide additional evidence for the necessity of incorporating radiation and cosmic rays in numerical galaxy formation simulations. Refining our understanding of galaxy formation models will, in turn, yield deeper insights into the puzzling early stages of galaxy formation.

\section{Acknowledgements}
TD and YY are supported by Isaac Newton Studentships from the Cambridge Trust. TD also acknowledges support from the UKRI Science and Technology Facilities Council (STFC) under grant number ST/V50659X/1. SMA acknowledges support from a Kavli Institute for Particle Astrophysics and Cosmology (KIPAC) Fellowship, from the NASA/DLR Stratospheric Observatory for Infrared Astronomy (SOFIA) under the 08\_0012 Program, and from the UKRI STFC under grant number ST/N000927/1. SMA also acknowledges visitor support from the Kavli Institute for Cosmology, Cambridge. DS acknowledges support from the STFC. The \textsc{Azahar} simulations studied here were generated with the new Cosma 8 HPC facility: this work used the DiRAC@Durham facility managed by the Institute for Computational Cosmology on behalf of the STFC DiRAC HPC Facility (www.dirac.ac.uk). The equipment was funded by BEIS capital funding via STFC capital grants ST/P002293/1, ST/R002371/1 and ST/S002502/1, Durham University and STFC operations grant ST/R000832/1. DiRAC is part of the National e-Infrastructure. Part of the \textsc{Azahar} simulations analysis employed the Sherlock cluster at Stanford: some of the computing for this project was performed on the Sherlock cluster. We would like to thank Stanford University and the Stanford Research Computing Center for providing computational resources and support that contributed to these research results.

\section{Data Availability}
\label{s_data_availability}
Post-processing scripts are made available upon reasonable request.

\bibliographystyle{mnras}
\bibliography{refs}

\label{lastpage}
\end{document}